\documentclass[submission, PhysLectNotes]{SciPost}

\hypersetup{
    colorlinks,
    linkcolor={red!50!black},
    citecolor={blue!50!black},
    urlcolor={blue!80!black}
}

\usepackage[bitstream-charter]{mathdesign}
\usepackage{amsmath}
\usepackage{wrapfig}
\DeclareSymbolFont{usualmathcal}{OMS}{cmsy}{m}{n}
\DeclareSymbolFontAlphabet{\mathcal}{usualmathcal}

\definecolor{cred}{RGB}{214, 39, 40}
\definecolor{cblue}{RGB}{31, 119, 180}
\definecolor{corange}{RGB}{255, 127, 14}

\definecolor{oldiagcolorB}{HTML}{A91E22} %ged
\definecolor{oldiagcolorA}{HTML}{A7A9AC} %gray
\definecolor{oldiagcolorC}{HTML}{1B75BC} %blue
\definecolor{oldiagcolorD}{HTML}{646567}

\newcommand{\Eq}[1]{Eq.$\:$(\ref{#1})}

\newcommand{\Fig}[1]{Fig.$\:$\ref{#1}}

\newcommand{\mur}{\mu_R^2}
\newcommand{\muf}{\mu_F^2}

\newcommand{\xip}{\xi_{p}}

\newcommand{\partsec}[2]{\frac{d \hat{\sigma}_{#2}^\mathrm{#1}}{ d \xi_p}}

\definecolor{basecolor}{HTML}{000066}

\begin{document}

\begin{center}{\Large \textbf{
Approximating missing higher-orders in transverse momentum distributions using resummations \\
}}\end{center}

\begin{center}
Niccolò Laurenti \textsuperscript{1},
Tanjona R. Rabemananjara \textsuperscript{2,3} and
Roy Stegeman \textsuperscript{1}
\end{center}

\begin{center}
{\bf 1} TIF Lab, Dipartimento di Fisica, Università degli Studi di Milano and INFN Sezione di Milano
\\
{\bf 2} Nikhef Theory Group, Science Park 105, 1098 XG Amsterdam, The Netherlands
\\
{\bf 3} Department of Physics and Astronomy, Vrije Universiteit, NL-1081 HV Amsterdam
\\
{\small \sf DIS2022: XXIX International Workshop on Deep-Inelastic Scattering and Related Subjects, Santiago de Compostela, Spain, May 2-6 2022.}
\end{center}

\begin{center}
\today
\end{center}

\section*{Abstract}
{\bf
We present a more reliable approach to approximate the unknown next-to-next-to-next-to-leading order (N3LO) 
transverse momentum distribution of colourless final states, namely the Higgs boson produced via gluon fusion 
and the lepton pair produced via Drell--Yan (DY) mechanism. The approximation we construct relies on the 
combination of various resummation formalisms -- namely threshold, small-pt and high energy resummations -- by
exploiting the singularity structure of the large logarithms in Mellin space. We show that for the case 
of Higgs boson production, the approximate N3LO transverse momentum distribution amounts to a correction 
of a few percent with respect to the NNLO result with a reduction in the scale dependence.
}

\vspace{1pt}
\noindent\rule{\textwidth}{1pt}
\tableofcontents\thispagestyle{fancy}
\noindent\rule{\textwidth}{1pt}
\vspace*{-20pt}

\section{Introduction}
\label{sec:intro}

In order to push forward precision and discovery physics at the LHC, the three pillars of QCD -- namely fixed-order
calculations, resummations, and PDFs -- need to be determined at the highest accuracy possible (sub one percent). 
While significant, much
work remains on the side of perturbative calculations, both in terms of fixed-order computations and resummations. Given that pushing the accuracy of the
fixed-order computations is a gigantic task, it is crucial to estimate the missing higher-order contributions to the best of our abilities. Currently, the most commonly used way of estimate theoretical uncertainties associated with missing
higher-order contributions is by varying the unphysical scales involved in the process according to a 
\emph{scale variation prescription}. On one hand, the use of scale variation to estimate the missing higher-order
uncertainties (MHOU) presents several advantages. First, due to the fact that the scale dependence of the strong
coupling and PDFs are universal, scale variations can be used to estimate theoretical uncertainties for any 
perturbative processes. Second, the constraint imposed by the renormalization group invariance ensures that as the
order of the perturbative calculations increases, the scale dependence decreases. Third, the estimation of MHOU resulting from scale variations produces smooth functions of the kinematics, accounting for the correlations in the
nearby regions of the phase space. On the other hand, the scale variation method has a number of caveats, chief among
which is the fact that it does not allow for a probabilistic interpretation. In addition, there is the ambiguity in defining the central scale around which the
variation should be performed and the ranges at which the scales are allowed to vary. But most importantly, scale
variation misses uncertainties associated to new singularities appearing at higher-orders but not present at lower-orders. Various approaches~\cite{Cacciari:2011ze, Bagnaschi:2014wea, Bonvini:2020xeo} have recently merged 
in order to address the shortcomings related to the scale variation method. However, most of these approaches are 
only applicable to particular types of processes and observables. In Ref.~\cite{Cacciari:2011ze} , Cacciari and 
Houdeau proposed a new approach of estimating MHOU using a Bayesian
model. In short, the method consists on adopting some assumptions on the progression of the perturbative expansion,
then based on the knowledge of the first few orders, one can infer on the \emph{hidden parameters} that are
assumed to bound the structure of the perturbative coefficients, allowing for an inference on the unknown subsequent
contributions. While this approach has proved to perform well for QCD observables at $e^+e^-$ 
colliders~\cite{Cacciari:2011ze}, its reliability when it comes to proton-proton collider observables is subject 
to question. In Ref.~\cite{Bonvini:2020xeo}, Bonvini built upon Cacciari-Houdeau's work to construct more general,
flexible, and robust models. The various models have been validated on various inclusive observables at the LHC.
However, none of these models can as of yet be used for differential observables as the correlations between the 
different regions are not accounted for.

A possible way to estimate (or rather approximate) the impacts of unknown higher-orders in perturbation theory is by using
the information provided by the resummed calculations. Given that the information on the various kinematic limits
that appear in fixed-order calculations are contained to all-order in resummed expressions, it should be possible
to consistently combine these various limits to approximate the subsequent unknown contributions. A similar
approximation was done decades ago in the context of the Higgs boson production from gluon fusion at the total
inclusive level~\cite{Ball:2013bra}. In the following, we provide a proof of concept on how to combine the different resummation formulae
to approximate the N3LO transverse momentum distributions for the Higgs boson productions in the infinite top mass limit. The eventual aim of the project is to extend the formalism to DY processes and use it to approximate MHOU in PDF fits. Throughout this
manuscript, the counting refers to the order at the integrated level, i.e. N3LO refers to $\mathcal{O}(\alpha_s^5)$
in the expansion of the transverse momentum distributions.

\section{Higgs boson production in HEFT}

To the present day, the inclusive and transverse momentum distributions of the Higgs boson produced via gluon
fusion are known to a fairly high precision. The total inclusive cross section with finite top-quark mass is
known up to N3LO~\cite{Mistlberger:2018etf}. Recently, the N3LO$\star$ transverse momentum distributions for 
the Higgs produced with an associated jet in the final state have become available~\cite{Caola:2015wna}. However, 
since these results are often obtained numerically, reading off the coefficients that are relevant for the 
comparison to the all-order computations is not practically feasible.

In the following sections, we present a formalism for the construction of the Higgs transverse momentum distribution
beyond NNLO by combining the information on the singularity structure at small-$N$ and large-$N$ which can be
predicted by the high-energy and threshold resummation formulae, respectively. Threshold resummation embodies to
all orders in $\alpha_s$ logarithms of the form $\ln N$ that drives the transverse momentum spectra in the limit
$N \to \infty$, while the $(N-1)^n$ behaviour for $N \to 1$ is fully determined to all orders by the small-$x$
resummation. Therefore, for a partonic cross section known up to N$^n$LO (i.e. $\mathcal{O}(\alpha_s^{n+2})$),
the approximate expression is constructed as a combination of fixed-order calculations and expansion from resummations: 
\begin{flalign}
	& \left[ \partsec{}{ab} \right]^{\text{N}^{n+1}\text{LO}} \hspace*{-1cm} (N,\xi_p)
	= \left[ \partsec{}{ab} \right]^{\text{N}^n\text{LO}} \hspace*{-0.7cm} (N,\xi_p)
	+ \partsec{}{ab}^{\hspace*{-0.3cm}(n+1)} \hspace*{-0.4cm} (N, \xi_p) \qquad \text{with} \qquad
	\partsec{}{ab}^{\hspace*{-0.3cm}(n)} = \partsec{}{ab}^{\hspace*{-0.3cm}\mathrm{he}, (n)}
	 + \partsec{}{ab}^{\hspace*{-0.3cm}\mathrm{th}, (n)}, \label{eq:approximation}
\end{flalign}
where $\xip \equiv p_T^2/M^2$ is a dimensionless variable with $M$ the
invariant mass of the produced Higgs boson. Notice that no matching function is introduced when combining
the two resummed cross sections. This means that the second part of \Eq{eq:approximation} is only valid if the
small-$N$ behaviour controlled by the high-energy contribution is not spoiled by the threshold component and
vice-versa.

In order to construct our approximate expression, let us first describe the large-$N$ approximation of the partonic
cross section in \Eq{eq:approximation}. The analytical expression of the threshold resummation for transverse
momentum distributions with colour singlet in the final state has been derived in~\cite{Forte:2021wxe}. By expanding 
the expression as a series in $\alpha_s$ one can extract the terms relevant for the approximation. However, the way
in which the logarithms of $N$ appear in the expansion does not correspond to the Mellin transform of the
$x$-space expressions which spoils the finite-$N$ behaviour. In addition, the resulting expressions display 
unphysical singularities. Indeed, as opposed to exhibiting poles at small-$N$, the expressions contain a logarithmic
branch cut at $N=0$. The correct singularity structure in the small-$N$ region of the resummed expression
can be restored by resorting to the $\Psi$-soft prescription in which the logarithms of $N$ are simply replaced
by the digamma functions as has been done for inclusive Higgs production in Ref.~\cite{Ball:2013bra}. Doing so yields the following resummed expression:
\begin{align}
	\hspace*{-0.25cm}
	\partsec{\mathrm{th}}{ab} \left( N, \xip, \frac{Q^2}{\mur}, \frac{Q^2}{\muf} \right)
	= \tilde{\mathcal{C}}_{ab} (N, \xip) \sum_{n=1}^{\infty} \alpha_s^n \sum_{k=0}^{2n}
	\tilde{g}_{n,k} \left( N, \xip, \frac{Q^2}{\mur}, \frac{Q^2}{\muf} \right) 
	\left( \psi_{0}(N) + \gamma_{E} \right)^k 
	\label{eq:theshold}
\end{align}
where $\tilde{\mathcal{C}}$ collects all the non-logarithmic dependence and $\tilde{g}_{n,k}$ are numerical
coefficients. We should emphasize that in the above equation the born-level cross section which contains
$\mathcal{O}(\alpha_s^2)$ contribution is included in the definition of the coefficient $\tilde{\mathcal{C}}$.

\begin{figure}[t]
	\centering
	\includegraphics[width=\linewidth]{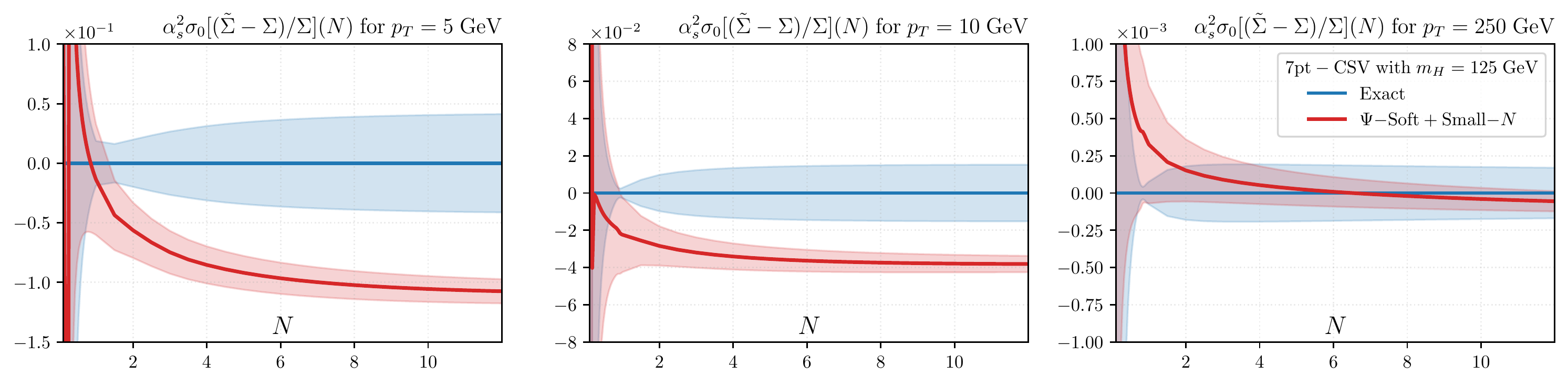}
	\caption{Plots comparing the relative difference between the exact and approximate solutions at NNLO for the 
		$gg$-channel at different values of $p_T$. The error bands have been computed using the $7$-point
		scale variation method. The codes used to generate the predictions both in momentum and Mellin
		space are publicly available~\cite{Rabemananjara_Stegeman_2021, Rabemananjara_Stegeman_2021_hptN3lo}.
	} \label{fig:nnlo}
\end{figure}

The leading logarithmic (LL$x$) high-energy resummation for the transverse momentum distribution of the
Higgs boson in HEFT has been derived in Ref.~\cite{Forte:2015gve}. The computations were performed by keeping 
the initial-state gluons off their mass-shell, $p_i^2 = \vert p_{T,i}\vert^2$ from which the \emph{impact
parameter} that defines the cross section is derived. For the $gg$-channel, for instance, the expanded 
resummed expression up to NNLO is written as:
\begin{align}
	\frac{d \hat{\sigma}^{\mathrm{he}}_{gg}}{d\xip} \left( N, \xip \right)
	= \sigma_{\mathrm{H},gg}^{\tiny \text{Born}} \left[ \left( \frac{\alpha_s}{\pi}
	\right) \frac{C_A}{N} \frac{2}{\xip} + \left( \frac{\alpha_s}{\pi} \right)^2
	\left( \frac{2 C_A}{N} \right)^2 \frac{\ln \xip}{\xip} \right]
	+ \mathcal{O} (\alpha_s^3) .
	\label{eq:high-energy}
\end{align}
In view of combining the high-energy approximation and the threshold approximation, one has to make sure
that the small-$N$ contributions vanish at moderately large-$N$. From \Eq{eq:high-energy} one
can see that not only the distribution always vanish in the large-$N$ limit but also the vanishing point
is located at the vicinity of $N \sim 1$ where effects from the threshold limit start to contribute.

In \Fig{fig:nnlo} we compare the approximation to the exact NNLO result. By focusing first on
the small-$N$ region ($N < 1$), it is apparent that the high-energy approximation reproduces
the fixed-order computations fairly well. Not only do the uncertainty bands of the two results overlap, but in all
cases the uncertainty bands of the exact results are contained in the approximation. Moving to the region
where $N > 1$, one notices that small discrepancies persist between the exact and the approximate results.

\begin{wrapfigure}{l}{0.5\textwidth}
	\begin{center}
		\includegraphics[width=0.5\textwidth]{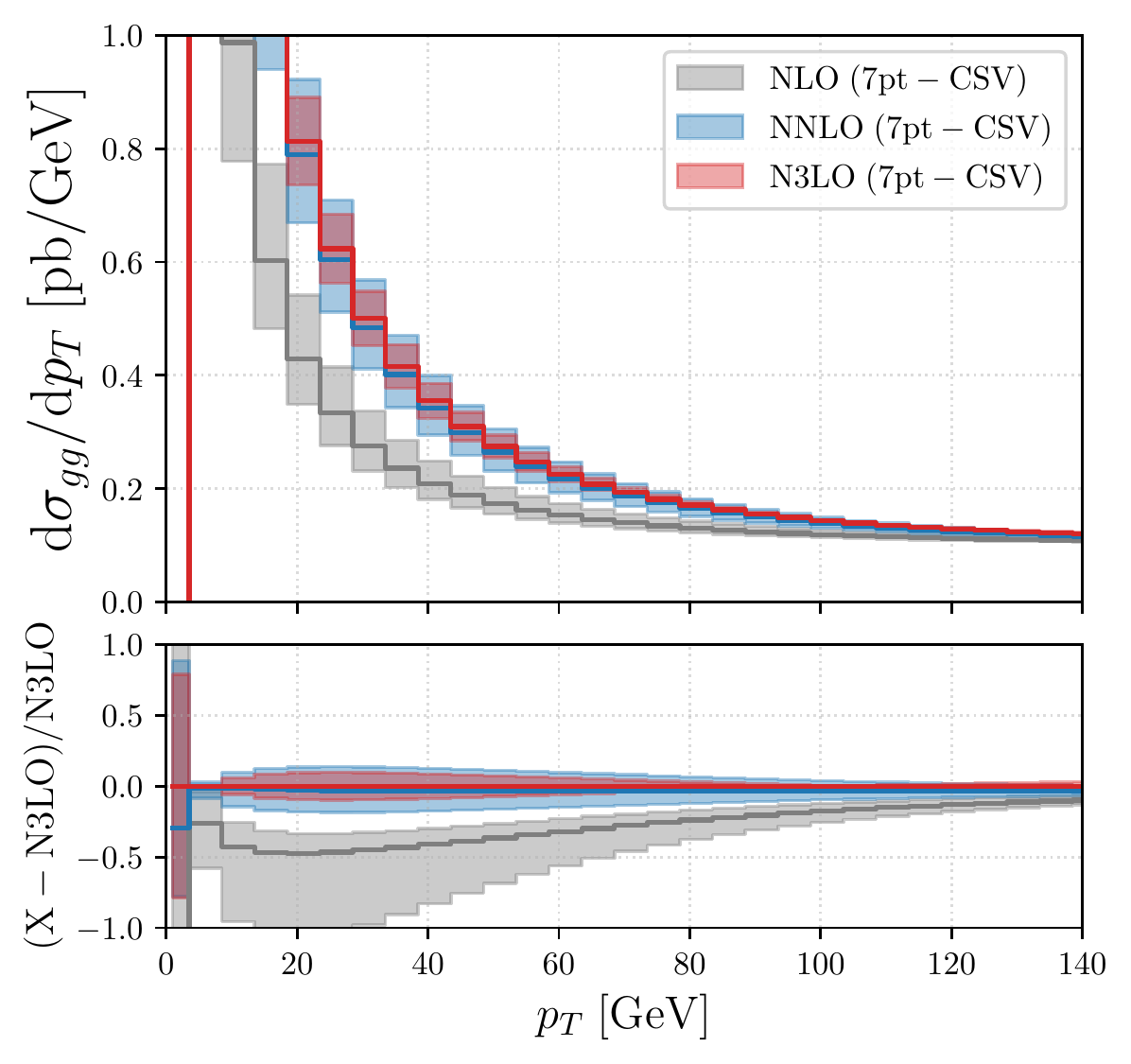}
	\end{center}
	\caption{Approximate N3LO higgs transverse momentum distribution in HEFT approximation at $\sqrt{s} = 14~\mathrm{TeV}$.}
	\label{fig:n3lo}
	\vspace*{-0.5cm}
\end{wrapfigure}

These discrepancies however reduces as the value of the transverse momentum increases. The approximation
would improve slightly if the contributions from the small-$p_T$ resummation were included. Nevertheless, omitting
the small-$p_T$ contributions in the HEFT is justified by the fact that it coincides with the high-energy
contributions at small-$p_T$ and large-$N$.

In \Fig{fig:n3lo} we show the approximate N3LO Higgs transverse momentum distribution at the hadronic
level. The inverse Mellin transform is computed using the contour deformation defined by the \emph{
Minimal Prescription} as described in~\cite{Rabemananjara:2020rvw}. For comparisons, both the exact NLO 
and NNLO are also included. As in the previous figures, the uncertainty bands have been computed using 
the $7$-point scale variation. One can see that the approximate N3LO transverse momentum distribution 
amounts to a correction of a few percent with respect to the NNLO result. As expected, the uncertainty 
band from the N3LO approximation is smaller compared to the one from NNLO with the former fully contained 
in the latter.\\

\section{Drell--Yan process}
The results presented here are for the Higgs production in the infinite top mass limit provides a relatively simple case study as a proof of concept for the methodology. However, the eventual aim is to provide a set of parton distribution functions (PDFs) accounting for MHOU approximated using resummations. For this reason it is important to apply the methodology also to DY processes as it provides valuable information about the proton structure.

The way in which the approximate N3LO expression is constructed for DY case is very similar to the Higgs
described above with the main difference that in the DY case the contributions from the small-$p_T$ resummation have
to be taken into account. While the analytical expressions of the small-$p_T$ and threshold resummation
formulae are available in the literature~\cite{Forte:2021wxe, Muselli:2017bad, Rabemananjara:2020rvw}, only the large-$b$ (or equivalently small-$p_T$) version is known for the
high-energy limit in Ref.~\cite{Marzani:2015oyb}, which extends the formalism for the differential cross-section for Higgs production as described in Ref.~\cite{Forte:2015gve}. Since the resummation formalism in the small-$p_T$ is known for DY processes, we are instead interested in the limit of large transverse momentum. This requires redoing the calculation for DY transverse momentum presented in Ref.~\cite{Marzani:2015oyb}.
It is finally worth noting that the derivation of the full expression is a 
complicated task given
the non-trivial relation between $\ln N$ and $\ln p_T$. 
% While development on this area is ongoing,
% results are already available for the combination of the small-$p_T$ and threshold contributions. The
% combination is done through a matching function as described in~\cite{Rabemananjara:2020rvw}. The results 
% in which we validate the procedure at NNLO are shown in \Fig{fig:n3lo} where we can see that the difference 
% between the exact and the combined resummed expressions is very small.

\vspace*{-0.5cm}
\section{Conclusions}
We explored the idea of using all-order computations to approximate contributions from missing
higher-orders. The combination of the various resummation formalisms were carried out in Mellin
space where one can fully study the singularity structure of the resummed expressions. Such an
approximation seem to yield reasonable predictions as attested by the partonic and hadronic
results. However, further work is required in order to apply the approximation also to DY processes in particular for the goal of using the approximation in PDF fits.

\vspace*{-0.35cm}
\paragraph{Acknowledgment:} The authors thank Stefano Forte for useful discussions on combined resummation.
T.R. and R. S are supported by the European Research 
Council under the European Union’s Horizon 2020 research and innovation Programme (grant agreement n.740006).
T. R. is also supported by an ASDI grant of The
Netherlands eScience Center.

\vspace*{-0.5cm}
\bibliography{bibliography.bib}

\nolinenumbers

\end{document}